\documentclass[fleqn,twoside]{article}
\usepackage{espcrc2}

\usepackage{graphicx}

\def\fun#1#2{\lower3.6pt\vbox{\baselineskip0pt\lineskip.9pt
  \ialign{$\mathsurround=0pt#1\hfil##\hfil$\crcr#2\crcr\sim\crcr}}}

\begin{document}

\title{High Energy Astroparticle Physics}

\author{G\"unter Sigl\address[oha]{APC (AstroParticules et Cosmologie),
   10, rue Alice Domon et Léonie Duquet, 75205 Paris Cedex 13, France and\\
Institut d'Astrophysique de Paris, 98bis Boulevard Arago,
 75014 Paris, France}}

\begin{abstract}
We give a brief (and highly incomplete) overview of the current
experimental and theoretical
status of high energy cosmic rays and their secondary $\gamma-$rays
and neutrinos. We focus on the role of large scale magnetic fields
and on multi-messenger aspects linking these three channels. We also
recall that the flavor composition of
neutrino fluxes from astrophysical sources contains information
on both the source conditions and neutrino physics.
\vspace{1pc}
\end{abstract}

\maketitle


\section{Introduction}

After almost 90 years of research, the origin of cosmic rays
is still an open question, with a degree of uncertainty
increasing with energy~\cite{crbook}. Above $\sim100\,$MeV the CR spectrum
exhibits little structure and is approximated
by broken power laws $\propto E^{-\gamma}$:
At the energy $E\simeq4\times 10^{15}\,$eV
called the ``knee'', the flux of particles per area, time, solid angle,
and energy steepens from a power law index $\gamma\simeq2.7$
to one of index $\simeq3.0$. The bulk of the CRs up to at least
that energy is believed to originate within the Milky Way Galaxy,
typically by shock acceleration in supernova remnants.
These objects have been seen in $\gamma-$rays up to $\sim10\,$TeV by
experiments such as HESS~\cite{hess} and MAGIC~\cite{Sidro:2006zq}, with a 
spectrum roughly scaling
as $E^{-2.2}$. This is consistent with $\gamma-$ray production in interactions
with the ambient gas of primary cosmic rays with a similar spectrum
at the source. The difference to the cosmic ray spectrum observed to
scale as $E^{-2.7}$ below the knee can be explained by diffusion in the
galactic magnetic fields.

Above the knee the spectrum continues with a further steepening to 
$\gamma\simeq3.3$
at $E\simeq4\times 10^{17}\,$eV, sometimes called the ``second knee''.
There are experimental indications that the chemical composition
changes from light, mostly protons, at the knee to domination by
iron and even heavier nuclei at the second knee~\cite{Hoerandel:2004gv}.
This is in fact expected in any scenario where acceleration and
propagation is due to magnetic fields whose effects only depend
on rigidity, the ratio of charge to rest mass, $Z/A$. This is true
as long as energy losses and interaction effects, which in general depend
on $Z$ and $A$ separately, are small, as is the case in the Galaxy, in
contrast to extra-galactic cosmic ray propagation at ultra-high energy.
Above the so called ``ankle'' or ``dip'' at $E\simeq5\times10^{18}\,$eV, the
spectrum flattens again to a power law of index $\gamma\simeq2.8$.
This latter feature is often interpreted as a cross over from a Galactic
component, which steepens because cosmic rays are not confined by
the galactic magnetic field any more or because Galactic sources
do not accelerate beyond the ankle, to a harder component of extragalactic
origin. However, the dip at $E\simeq5\times10^{18}\,$eV could also be
explained by pair production by extra-galactic protons,
if the extra-galactic component already starts to dominate below
the ankle, for example, around the second-knee~\cite{Aloisio:2006wv}
at a few times $10^{17}\,$eV. This requires a relatively steep injection
spectrum $\propto E^{-2.6-2.7}$. Below a few times $10^{17}\,$eV
this extra-galactic component would become unobservable at Earth
due to diffusion in extra-galactic magnetic fields (EGMF)~\cite{Lemoine:2004uw}.
In addition, the effective volume-averaged injection spectrum has to become 
flatter somewhere below $\sim10^{18}\,$eV in order to avoid excessive
power going into cosmic rays and to avoid overproduction
of GeV--TeV $\gamma-$rays from $pp$ interactions with the ambient gas.

The low cross-over scenario also requires the dominance of protons around
the dip. Theoretically, this can be achieved either because preferentially 
protons are accelerated or because extended EGMF lead to strong
photo-spallation during propagation~\cite{Sigl:2005md}.
Experimentally, above $\simeq10^{17}\,$eV the chemical composition is basically unknown~\cite{Watson:2004ew}. Around $10^{18}\,$eV the situation is
particularly inconclusive as HiRes~\cite{Abbasi:2004nz} and HiRes-MIA~\cite{Abu-Zayyad:2000ay} data suggest a light (proton dominated) composition, whereas other experiments indicate a heavy composition~\cite{Hoerandel:2004gv}. In any case, the cosmic
ray flux should be extra-galactic at least above the ankle, where a
galactic origin would predict an anisotropy toward the
galactic plane because galactic magnetic fields can no longer
isotropize the cosmic rays. No such anisotropy is seen.
There are also experimental
indications for a chemical composition becoming again lighter above
the ankle, although a significant heavy component is not
excluded and the inferred chemical composition above
$\sim10^{18}\,$eV is sensitive to the model of air shower interactions
and consequently uncertain presently~\cite{Watson:2004ew}.
In addition, should a substantial heavy composition
be experimentally observed up to the highest energies, some sources
would have to be surprisingly nearby, within a few Mpc, otherwise only
low mass spallation products would survive propagation~\cite{Harari:2006uy}.
In the following we will restrict our discussion on extra-galactic
ultra-high energy cosmic rays (UHECRs).

No conclusive picture for the nature and distribution of the sources
emerges yet naturally from the data~\cite{Cronin:2004ye}: Arrival directions are approximately isotropic~\cite{bm}, suggesting a large number of weak or distant
sources. But there are also indications which point more towards a small number of
local and therefore bright sources, especially at the highest energies:
First, the AGASA ground array claimed statistically significant multi-plets of
events from the same directions within a few degrees~\cite{Shinozaki:2006kk,bm},
although this is controversial~\cite{Finley:2003ur} and has not been seen so far
by other experiments such as the fluorescence experiment HiRes~\cite{finley}.
The spectrum of this clustered component is $\propto E^{-1.8}$ and thus
much harder than the total spectrum~\cite{Shinozaki:2006kk}.
Second, nucleons above $\simeq70\,$EeV suffer heavy energy losses due to
photo-pion production on the cosmic microwave background (CMB)
--- the Greisen-Zatsepin-Kuzmin (GZK) effect~\cite{gzk} ---
which limits the distance to possible sources to less than
$\simeq100\,$Mpc~\cite{stecker}. This predicts a ``GZK cutoff'', a
drop in the spectrum, whose strength depends on the source distribution
and may even depend on the
part of the sky one is looking at: The ``cutoff'' could be mitigated
in the northern hemisphere where more nearby accelerators related
to the local supercluster can be expected. Apart from the SUGAR array
which was active from 1968 until 1979 in Australia, all UHECR detectors
completed up to the present were situated in the northern hemisphere.
Nevertheless the situation is unclear even there: Whereas a ``cut-off''
is consistent with the few events above $10^{20}\,$eV recorded
by the fluorescence detector HiRes~\cite{Bergman:2006vt} and with the first
data release of the Pierre Auger observatory~\cite{Kampert:2006ec},
there is a tension with the 11 events above $10^{20}\,$eV detected
by the AGASA ground
array~\cite{agasa}. Still, this could be a combination of statistical
and systematic effects~\cite{DeMarco:2003ig}, especially given the recent
downward revision of the energy normalization in AGASA~\cite{teshima}.
The solution of this problem will have to await more analysis and
more statistics and, in particular, the completion of the Pierre Auger 
project~\cite{Kampert:2006ec}
which combines the two complementary detection techniques
adopted by the aforementioned experiments and whose southern site
is currently in construction in Argentina.
Finally, about 1\% of the HiRes steroa events around $10^{19}\,$eV
seem to correlate with active galaxies of the BL Lac on a scale of
$\sim0.6^\circ$, with a significance of
$\sim10^{-4}$~\cite{Abbasi:2005qy,Gorbunov:2004bs}. Due to deflection of UHECR
in the galactic magnetic field this, however, would have to be
neutral primaries that cannot be created in the necessary quantities
over the distances involved.

\section{Role of large scale magnetic fields}

The hunt for UHECR sources is further complicated by the presence
of large scale cosmic magnetic fields which may significantly deflect
charged cosmic rays even at the highest energies, in particular if
sources correlate with high magnetic field regions such as galaxy
clusters. A major issue in UHECR 
propagation studies is, therefore, the strength and distribution of EGMF.
Galaxy clusters harbor magnetic fields of $\mu$G
strength, but it is poorly known how quickly these fields fall
off with increasing distance from the cluster center. The current data
indicate that $\mu$G strength magnetic fields extend to at least $\sim 1$
Mpc~\cite{clarke} and possibly to larger distances~\cite{johnston,Govoni:2006fs}.
Beyond $\simeq1\,$Mpc from a cluster core, however, probing the
magnetic fields becomes extremely difficult because the Faraday
Rotation Measure loses sensitivity in low density
regions. Furthermore, the intracluster magnetic field topology is
also poorly known, although the situation will likely improve
in the future, for example with the advent of powerful radio astronomical
instruments such as the square kilometer array.

One possibility in the meantime is to adopt large scale structure simulations (LSS) which include magnetic fields. In Ref.~\cite{lss-protons}, the authors
use magnetic fields derived from a cosmological
LSS with magnetic fields generated at the shocks that
form during LSS formation, whereas in Ref.~\cite{sme-proc} and
Ref.~\cite{dolag} fields of ``primordial'' origin have been
considered. While the different models for initial magnetic seed fields
produce different large scale magnetic field distributions and,
therefore, lead to different predictions for UHECR deflection,
there is still a significant discrepancy between
Ref.~\cite{lss-protons,sme-proc} and  Ref.~\cite{dolag}, hinting
that other technical reasons may play a role here. In the more
extended fields from the simulations
of Refs.~\cite{lss-protons,sme-proc} deflection of protons up to
$10^{20}\,$eV can be up to tens of degrees, whereas deflections
in the simulations of Ref.~\cite{dolag} are typically below a degree.
Assuming the EGMF correlates with the infrared luminosity density,
Ref.~\cite{Elyiv:2006fv} recently found results closer to 
Refs.~\cite{lss-protons,sme-proc} than Ref.~\cite{dolag}.

We recall that since acceleration is rigidity dependent, at the
acceleration sites the highest energy cosmic ray flux is likely
dominated by heavy nuclei. If this is indeed the case, it is interesting
to point out that even in the EGMF scenario of Ref.~\cite{dolag},
deflections could be considerable. In
contrast to the contribution of our Galaxy to deflection which can be
of comparable size but may be corrected for within sufficiently
detailed models of the galactic field, the extra-galactic contribution
would be stochastic. Statistical methods are therefore likely to be
necessary to learn about UHECR source distributions and
characteristics as well as EGMF. For example, a suppressed UHECR arrival 
direction auto-correlation function at degree scales, rather than
pointing to a high source density, could be a signature of
extended EGMF~\cite{lss-protons}.

Finally, EGMF can considerably increase the path-length of UHECR
propagation and thus modify spectra, especially from individual sources,
as well as the chemical composition observed at Earth~\cite{spec-comp}.

\section{Multi-messenger approach: Secondary gamma-rays and neutrinos
and their flavor composition}

\begin{figure}[h!]
\includegraphics*[width=0.48\textwidth]{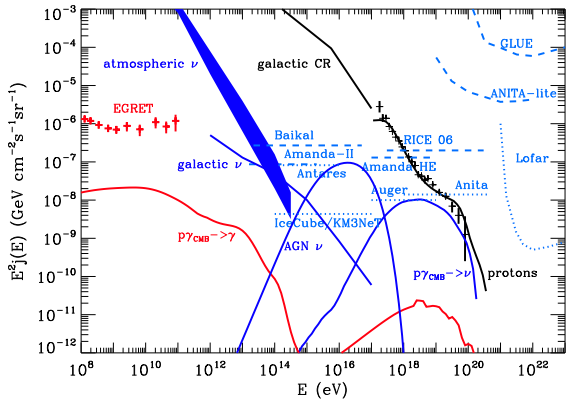}
\vskip-20pt
\caption{
Model fluxes compared to experimental
data, limits and sensitivities. Primary cosmic ray fluxes (data and a model,
see text) are shown in black, the secondary  $\gamma-$ray flux expected from proton interactions with the CMB and infrared background
in red and the "guaranteed" neutrino fluxes per neutrino flavor in blue: 
atmospheric neutrinos, galactic neutrinos resulting from cosmic ray interactions with matter in our Galaxy~\cite{Candia:2003ay}, and "GZK'' neutrinos resulting 
from cosmic ray interaction with the CMB and infrared background. The GZK neutrino fluxes depend on the distribution of the (unknown) primary cosmic ray sources for which we assumed active galactic nuclei (AGNs) above
$10^{17}\,$eV, using our public CRPropa code~\cite{crpropa,Armengaud:2006fx}.
Cosmic ray interactions within these sources can also produce neutrinos
for which one example is given (AGN $\nu$)~\cite{Semikoz:2003wv}.
The flux of atmospheric neutrinos has been measured by underground detectors and AMANDA. The dashed and dotted
blue lines are existing upper limits and future sensitivities to diffuse neutrino fluxes from various experiments, respectively~\cite{nu_review}, assuming the Standard Model neutrino-nucleon
cross section extrapolated to the relevant energies. The maximum possible neutrino flux would be given by horizontally extrapolating the diffuse $\gamma-$ray background observed by EGRET~\cite{Strong:2003ex}.}
\label{fig1}
\end{figure}

The physics and astrophysics of UHECRs are also intimately linked with
the emerging field of neutrino astronomy~\cite{nu_review} as well as
with the already well established field of $\gamma-$ray 
astronomy~\cite{gammarev}. Indeed, all scenarios of cosmic ray origin,
from the galactic scale~\cite{Kistler:2006hp} to 
top-down~\cite{Bhattacharjee:1998qc} and Z-burst
models at the highest energies~\cite{Semikoz:2003wv}, are severely
constrained by neutrino and $\gamma-$ray observations and limits.
This has, for example, important consequences for theoretical
predictions of diffuse fluxes of extragalactic neutrinos above about a TeV
whose detection is a major goal of next-generation
neutrino telescopes: If these neutrinos are
produced as secondaries of protons accelerated in astrophysical
sources and if these protons leave the sources and
contribute to the UHECR flux observed, then
the energy content in the neutrino flux can not be higher
than the one in UHECRs, leading to the so called Waxman-Bahcall
bound for transparent sources with soft acceleration
spectra~\cite{wb-bound,mpr}.
If one of these assumptions does not apply, such as for acceleration
sources with injection spectra harder than $E^{-2}$ and/or opaque
to nucleons, or if much fewer nucleons than $\gamma-$rays and neutrinos
are produced, such as in top-down scenarios,
the Waxman-Bahcall bound does not apply, but the neutrino
flux is still constrained by the observed diffuse $\gamma-$ray
flux in the GeV range.

Fig.~\ref{fig1} provides a sketch of "realistic'' cosmic
ray, $\gamma-$ray, and neutrino flux predictions in comparison
with experimental observations, limits, and sensitivities. It
shows a theoretical scenario in which
extra-galactic cosmic ray sources roughly evolving as quasars
inject a spectrum $\propto E^{-2.6}$ of dominantly protons down to 
$\sim10^{17}\,$eV where a cross-over to galactic cosmic rays occurs~\cite{Aloisio:2006wv}. The "cosmogenic'' neutrino flux
produced by protons interacting with the low energy photon
background considerably depends on these assumptions which
can thus be used to test them~\cite{Stanev:2006su}.

Apart from cosmogenic neutrinos produced during propagation
of UHECR, neutrinos can also be produced within astrophysical
sources such as AGNs (see Fig.~\ref{fig1}) or $\gamma-$ray bursts.
In the absence of matter effects, a source at cosmological
distances injecting neutrino fluxes with a flavor ratio
$\propto w_\beta$, $\beta=e,\mu,\tau$, leads to a
flavor mixture $\phi_\alpha\propto
\sum_{\beta,i}w_\beta|U_{\alpha i}|^2|U_{\beta i}|^2$
observed at Earth, where $U_{\alpha i}$ is the mixing matrix and $i$
labels mass eigenstates. Therefore, if both pions and muons
decay before loosing energy around the source, $w_e:w_\mu:w_\tau\simeq1:2:0$
and thus $\phi_e:\phi_\mu:\phi_\tau\simeq1:1:1$. At high energies
the meson and muon energy loss time $t_{\rm loss}(E)$ becomes shorter
than their decay time $E\tau/m$, and the neutrino spectrum will be
suppressed by a factor $\simeq mt_{\rm loss}(E)/(\tau E)$ compared to
primary interaction rates.
For hadronic cooling, $t_{\rm loss}\sim$const, whereas for radiative
cooling at the highest energies, $t_{\rm loss}(E)\propto E^{-1}$,
resulting in a steepening of the neutrino spectrum by a factor $E^{-1}$
and $E^{-2}$, respectively~\cite{Ando:2005xi}. In addition, at a given energy,
charged pions decay about hundred times faster than muons.
There can thus be an energy range at which pions but not muons
decay before loosing energy such that $w_e:w_\mu:w_\tau\simeq0:1:0$ and thus
$\phi_e:\phi_\mu:\phi_\tau\simeq1:2:2$. Also, $pp$ interactions produce
both pions of both charges and thus give a higher fraction of $\bar{\nu_e}$
compared to $p\gamma$ interactions.
The observed flavor ratios can thus depend on energy and carry
information on the source conditions~\cite{Kashti:2005qa}, but also
about the mixing matrix itself~\cite{Serpico:2005bs}.

Finally, flavor
ratios can probe new physics, such as neutrino decay and quantum
decoherence~\cite{Hooper:2005ea}: If all but the lightest mass
eigenstate $j$ decay before reaching
the observer, the flux of flavor $\alpha$ observed at Earth would be
$\propto|U_{\alpha j}|^2$, independent of the flavor ratio at the source.
For $j=1$ (normal mass hierarchy) this gives
$\phi_e:\phi_\mu:\phi_\tau\simeq6:1:1$, whereas for $j=3$ (inverted
mass hierarchy) one has $\phi_e:\phi_\mu:\phi_\tau\simeq0:1:1$, which
should be easy to distinguish from the normal case. This would
allow to probe lifetimes of the order
$\tau/m\sim300(E/PeV)^{-1}\,$s/eV, which could improve on current limits.
Quantum decoherence
would predict $\phi_e:\phi_\mu:\phi_\tau=1:1:1$, independent of source
flavor ratios.

\begin{figure}[h!]
\includegraphics*[width=0.48\textwidth]{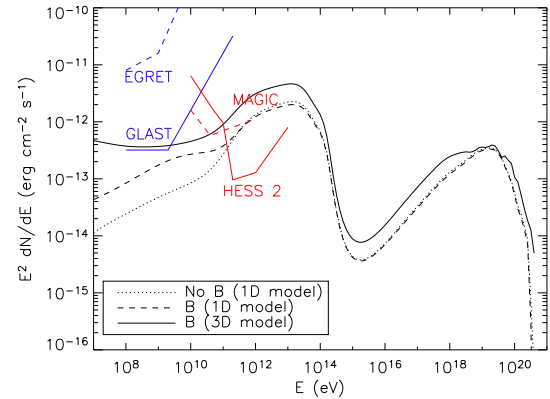}
\vskip-20pt
\caption{Differential $\gamma$-ray fluxes from electromagnetic cascades
including synchrotron radiation, initiated by
photo-pion and pair production by protons injected with an $E^{-2.7}$ spectrum
(not shown) by a source at distance $d=20\,$Mpc, from Ref.~\cite{armengaud}.
We assume the source contributes a 
fraction $\simeq 0.2$ to the total UHECR flux, corresponding to a proton luminosity $\simeq 4 \times 10^{42}\,$erg s$^{-1}$ above
$10^{19}\,$eV. A structured magnetic field of 0.1--1$\mu$G extends a few Mpc around the source in case of the 1D and 3D simulations which take into
account synchrotron radiation of electrons and positrons. The 1D model
neglects proton deflection whereas the 3D simulation follows 3-dimensional
proton trajectories. The latter case implies that the
fluxes shown here would be extended over $\sim5^\circ(20\,{\rm Mpc}/d)$
on the sky.
The fluxes are comparable to sensitivities of space-based $\gamma-$ray
(blue) and ground-based imaging air \v{C}erenkov detectors (red).}
\label{fig2}
\end{figure}

Three-dimensional propagation in structured large scale magnetic fields
also has considerable influence on secondary $\gamma-$ray and neutrino
fluxes. Fig.~\ref{fig2} demonstrates how magnetic fields of
$\mu$G strength surrounding a UHECR source, for example in
a galaxy cluster, can influence the secondary GeV-TeV $\gamma-$ray fluxes 
produced by electromagnetic cascades initiated by UHECR interactions
with the CMB and infrared background. This is the result of simulations
with our public code CRPropa~\cite{crpropa}, discussed in
Ref.~\cite{Armengaud:2006fx,armengaud}. For the
steep injection spectrum $\propto E^{-2.7}$ assumed in Fig.~\ref{fig2},
the photon flux below a TeV is dominated by synchrotron radiation from
electron/positron pairs produced by protons around the ankle. As a
consequence, it depends considerably on strength and extension of
EGMF around the source.

\bibliographystyle{aipprocl} 

\end{document}